\documentclass{article}
\usepackage{epsfig}

\begin{document}

\begin{titlepage}
\title{\bf Full $\bf{\it O(\bf \alpha_S)}$ Evaluation of $\;\bf b\rightarrow s \bf \gamma$
Transverse Momentum Distribution}
\author{Ugo Aglietti
\\
Dipartimento di Fisica, Universit\'a di Roma "La Sapienza", \\
INFN Sezione di Roma,\\ Piazzale A. Moro, Rome, Italy
\\
\\
Roberto Sghedoni
\\
Dipartimento di Fisica,  Universit\'a di Parma,\\
INFN Gruppo Collegato  di Parma,\\
Viale delle Scienze, Campus Sud, 43100 Parma, Italy
\\
\\
Luca Trentadue
\\
Dipartimento di Fisica, Universit\'a di Parma,\\
INFN Gruppo Collegato di Parma,\\
Viale delle Scienze, Campus Sud, 43100 Parma, Italy
\\
and
\\
Theory Division, CERN, CH-1211 Geneva 23, Switzerland}

\date{}
\maketitle

\vspace*{2.0cm}
\begin{abstract}
{\normalsize \noindent The full $O(\alpha_S)$ transverse momentum
distribution for the $b \rightarrow s \gamma$ decay is computed.
Results are presented in analytic form.
An improved expression for the coefficient function taking
into account subleading operators is given and an exact
expression for the remainder function associated with the leading
operator $\hat{\mathcal{O}}_7$ is also derived.}
\end{abstract}

\begin{picture}(5,2)(-330,-485)
\put(1.0,-5){DFUP-2003-25} \put(1.0,-15.2){CERN-TH/2003-255}
\put(1.0,-25.4){Rome1-1361/03}
\end{picture}
\vfill
\vspace{3cm}
\leftline{\hspace{1.2cm} October 29th 2003}
\thispagestyle{empty}

\end{titlepage}
%
%

\section{Introduction}
\setcounter{page}{1}
Perturbation theory, i.e. the expansion in powers of $\alpha_S$,
has been applied to describe decays of the beauty quark since its
discovery. While the expansion parameter  $\alpha_S(m_B)\sim
0.21$, being  reasonably small, allows one to have confidence in the
computations, it is difficult to directly compare the perturbative
approach with the experimental data. As is well known, decay rates do
not make good quantities to be compared with the data, because they
are proportional to the fifth power of the beauty quark mass, a
poorly known parameter,
\begin{equation}
\Gamma \propto m_b^5
\end{equation}
and because they involve in principle unknown CKM matrix elements
such as $V_{cb},~ V_{ub},~V_{ts},$ etc..
By taking ratios of different widths, one can cancel the $m_b^5$ dependence
in the observables, and, eventually, also the dependence
on the CKM matrix elements.
A rather good theoretical quantity is represented, for instance, by  the
semileptonic branching ratio:
\begin{equation}\label{slbf}
B_{SL}=\frac{\Gamma_{SL}}{\Gamma_{TOT}},
\end{equation}
which turns out to be marginally in agreement with
present data \cite{altpet}.
Inclusive quantities, $B_{inclusive}$,  such as (\ref{slbf}), have a
perturbative series that involves numerical coefficients $c_n$ of the form:
\begin{equation}
B_{inclusive}~=~ \sum_{n=0}^{\infty} c_n~ \alpha_S^n(m_B).
\end{equation}
In less inclusive quantities, additional dynamical effects
appear, due to the kinematical restrictions on the final particles,
and the use of perturbation theory is, in general, less justified.
In semi-inclusive quantities, $B_{semi-inclusive}$, such as threshold and transverse momentum $p_t$
distributions, the perturbative series contains large infrared
lo\-ga\-rithms in addition to the coefficients $c_n$; they may be expanded
as a perturbative series of the form:
\begin{equation}
B_{semi-inclusive}~=~
\sum_{n=0}^{\infty}\sum_{k=0}^{2n} c_{n,k} ~\alpha_S^n~ \log^k x,
\end{equation}
where $x$ represents the characteristic scale of the process
as the energy or the transverse momentum.
Resummation of such enhanced terms to any order in $\alpha_S$ can be
performed in various approximations.
\\
The simplest one, the leading logarithmic approximation,
involves picking up only the terms having two powers
of the logarithm for each power of the coupling, i.e. $k=2n$.
In the double-logarithmic approximation each parton is dressed
with a cloud of soft and collinear gluons. Further, more refined approximations involve
smaller numbers of logarithms for each power of $\alpha_S$, i.e. $k=2n-1, 2n-2,\dots$.
\\
In the last years, considerable effort has been devoted to the
study of various spectra in $B$ decays in the endpoint region, in
the framework of resummed perturbation theory.
\\
In order to verify the ability of the resummed perturbation theory
to describe $B$ decays in a different dynamical situation, we considered,
in a previous note \cite{noi}, $p_t$-distributions
describing that of the $s$ quark with respect to the photon direction, in the $b$ rest frame.
\\
In this work, \cite{noi}, the following issues have been considered:
the resummed $p_t$-distribution in the $b \rightarrow
s\gamma$  decay is evaluated and both perturbative and
non-perturbative sources of transverse momentum contributions
discussed. The general theoretical framework for the
evaluation of the corresponding matrix element  defined
and the strategy to evaluate leading and next-to-leading
perturbative contributions is outlined, by introducing
a method to treat the radiative corrections and their summation in
a improved perturbative formula.  The comparison of the transverse
momentum distribution singularity structure with the more
widely-known threshold case is also presented.
\\
The chosen quantity  manifests a clear advantage from a
phenomenological point of view since, as discussed in \cite{noi},
it depends only on the photon momentum in the process. Thanks to the
straightforward and direct kinematics, the transverse momentum turns
out to be a particularly simple variable to use to discuss the
singularity structure of the perturbative expansion.  The case of
a possible effective theory within which to factorize these singularities
can, for the transverse momentum, be considered as well.
\\
The general formula representing the
complete perturbative expression for a the resummed distribution is
given by the formula
\begin{equation}
D(x) = K(\alpha_S) \Sigma(x;\alpha_S) + R(x;\alpha_S).
\end{equation}
The results, already presented
in \cite{noi}, did concern the universal process-independent function
$\Sigma(x;\alpha_S)$, resumming the infrared logarithms in
exponentiated form.
\\
Here the general perturbative expression for the whole distribution
will be concisely recalled  and
the new entries represented by the coefficient function $K(\alpha_S)$ and
by the remainder function $R(x)$ will be evaluated. Both $K(\alpha_S)$ and $R(x)$ are process-dependent
and require an explicit evaluation of  Feynman diagrams.
\\
Resummation of large infrared logarithms in $b$ decays have been studied
in great detail in recent years. This scheme is justified by the fact that
the double logarithm appearing to order $\alpha_S$ can become rather large
(with respect to 1 coming from the tree level):
\begin{equation}
-\frac{\alpha_S C_F}{4\pi} \log^2\frac{p_t^2}{m_b^2} \sim -0.7
\end{equation}
if we push the transverse momentum to such small values as
$p_t~\sim~\Lambda_{QCD}~=~300$ MeV. The single logarithm can also become
rather large, having a large numerical coefficient:
\begin{equation}
-\frac{5\alpha_S C_F}{4\pi} \log\frac{p_t^2}{m_b^2} \sim 0.6.
\end{equation}
The purpose of resumming classes of such terms therefore seems quite justified.
If we consider running coupling effects, i.e. if the (frozen) coupling
evaluated at the hard scale $Q~=~m_B~=~5.2$ GeV is replaced by the coupling
evaluated at the gluon transverse momentum,
\begin{equation}
\alpha_S(m_b) \rightarrow \alpha_S(p_t)=0.45~~~~~~{\rm{for}}~~~~~p_t~=~1~\rm{GeV},
\end{equation}
the logarithmic terms have sizes of order:
\begin{equation}
-\frac{\alpha_S(p_t) C_F}{4\pi} \log^2\frac{p_t^2}{m_b^2} \sim -0.5
\end{equation}
and
\begin{equation}
-\frac{5\alpha_S C_F}{4\pi} \log\frac{p_t^2}{m_b^2} \sim 0.8.
\end{equation}
The main difference with respect to resummation in $Z^0$ decays is
a hard scale smaller by over an order of magnitude, i.e. a coupling larger by
a factor 2 and infrared logarithms smaller by a factor 3.

\section{The effective hamiltonian for the decay $b\rightarrow
s\gamma$}\label{sechamiltonian}
\begin{figure}[h]
\begin{center}
\mbox{\epsfig{file=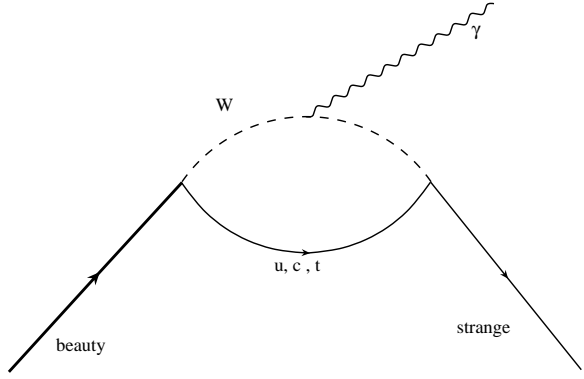, height=6cm}} \vskip -10mm
\caption{\label{vertice}Vertex for $b\rightarrow s\gamma$ in the Standard Model}
\end{center}
\end{figure}

The decay $b\rightarrow s\gamma$ is
loop-mediated in the Standard Model and offers stringent tests
of the latter as well as a way to extract CKM matrix elements.
The relevant diagrams involve a loop with a virtual
$W$ and an up-type quark ($u,c$ or $t$); the external
photon can be emitted from the internal lines and from the
external lines of the $b$ or $s$ quark  (see fig. \ref{vertice}).
QCD radiative corrections are affected by large logarithms of the form
\begin{equation}\label{logrisommati}
\alpha_S^n \ \log^k{m_W\over m_b} \ \ \ \ \ {\rm with}~~~0\le k\le n
\end{equation}
as well as logarithms of $m_t/m_W$. Since the energies involved in the
process are much smaller than the $W$ or $t$ mass, it is possible to
integrate out these fields by means of an operator product expansion
and write an effective low-energy hamiltonian of the form:
\begin{equation}\label{hamiltoniana}
{\cal H}_{eff}(x) = {G_F \over \sqrt{2}} \ V^*_{ts}V_{tb} \
\sum_{j=1}^8 C_j(\mu_b) \ \hat{\cal O}_j(x;\mu_b).
\end{equation}
With a factorization scale $\mu_b=O(m_b)$,
the long-distance effects --- both perturbative and non-perturbative ---
are factorized in the matrix elements of the operators $\hat{\cal O}_j$,
while the short-distance effects are contained in the
coefficient functions $C_j(\mu_b)$, calculable in perturbation theory.
In particular, the large logarithms in (\ref{logrisommati})
are included into the coefficient functions and can
be resummed with standard renormalization group techniques.
\\
A suitable basis for the operators $\hat{\cal O}_j$ is given by
six four-quark operators, $\hat{\cal O}_1$-$\hat{\cal O}_6$ and by
the penguin operators $\hat{\cal O}_7$, $\hat{\cal O}_8$
\cite{gsw1,gsw2}:
\begin{eqnarray}\label{base}
\hat{\cal O}_1 &=& (\overline{c}_{L,\beta} \gamma_\mu
b_{L,\alpha})(\overline{s}_{L,\alpha} \gamma_\mu
c_{L,\beta})\nonumber\\
\hat{\cal O}_2 &=& (\overline{c}_{L,\alpha} \gamma_\mu
b_{L,\alpha})(\overline{s}_{L,\beta} \gamma_\mu
c_{L,\beta})\nonumber\\
\hat{\cal O}_3 &=& (\overline{s}_{L,\alpha} \gamma_\mu
b_{L,\alpha})(\sum_q \overline{q}_{L,\beta} \gamma_\mu
q_{L,\beta})\nonumber\\
\hat{\cal O}_4 &=& (\overline{s}_{L,\alpha} \gamma_\mu
b_{L,\beta})(\sum_q \overline{q}_{L,\beta} \gamma_\mu
q_{L,\alpha})\nonumber\\
\hat{\cal O}_5 &=& (\overline{s}_{L,\alpha} \gamma_\mu
b_{L,\alpha})(\sum_q \overline{q}_{R,\beta} \gamma_\mu
q_{R,\beta})\nonumber\\
\hat{\cal O}_6 &=& (\overline{s}_{L,\alpha} \gamma_\mu
b_{L,\beta})(\sum_q \overline{q}_{R,\beta} \gamma_\mu
q_{R,\alpha})\nonumber\\
\hat{\cal O}_7 &=& {e\over 16\pi^2}m_{b,\overline{MS}}(\mu_b)
\overline{s}_{L,\alpha}\sigma^{\mu\nu}b_{R,\alpha}F_{\mu\nu}\nonumber\\
\hat{\cal O}_8 &=& {g\over 16\pi^2}m_{b,\overline{MS}}(\mu_b)
\overline{s}_{L,\alpha}\sigma^{\mu\nu}T^a_{\alpha\beta}b_{R,\alpha}G^a_{\mu\nu},\nonumber\\
\end{eqnarray}
where $m_{b,\overline{MS}}(\mu_b)$ is the $b$ mass in the $\overline{MS}$ scheme, evaluated at
$\mu_b$ and $q=u,d,s,c$ or $b$.
\\
The dimension of these operators is six: higher-dimension operators
have coefficients suppressed by inverse powers of the masses of the integrated
particles ($t$ and $W$) and do not contribute in first
approximation.
\\
The calculation of the QCD corrections to the coefficients functions has been carried out in
\cite{misiak1} with leading logarithmic accuracy and in
\cite{misiak2} at next-to-leading level in the $\overline{MS}$
scheme.
\\
Let us now consider the evaluation of the matrix elements of the effective
hamiltonian between quark states.
Only the magnetic penguin operator $\hat{\cal O}_7$ contributes in lowest order
with a rate:\footnote{
$\Gamma_0$ contains in principle
$m_{b,\overline{MS}}^{2}\left( \mu_{b}\right)$
since $\mu_b$ is the renormalization point of $\hat{\cal O}_7$.
As is well known, the renormalization point is arbitrary: we decided to fix it
to $m_b$ in the running mass, as usually done in the literature.
}
\begin{equation}\label{born}
\Gamma _0\simeq \frac{\alpha _{em}}{\pi }\frac{G_{F}^2\,m_{b}^{3}m_{b,
\overline{MS}}^{2}\left( m_{b}\right) \,|V_{tb}V_{ts}^{\ast
}|^{2}}{32\pi ^{3}}C_{7}^{2}\left( \mu_{b}\right),
\end{equation}
where $m_b$ is the pole mass of the $b$ quark.
\\
Radiative QCD corrections involve gluon brehmsstrahlung. The
operator $\hat{\cal O}_7$ is affected by infrared singularities
for the emission of a soft or a collinear gluon; the remaining
operators $\hat{\cal O}_1$-$\hat{\cal O}_6$ have infrared-finite
matrix elements. This implies that QCD corrections to the operator
$\hat{\cal O}_7$ only are logarithmically enhanced for $p_t\ll
m_b$ \footnote{ The operator $\hat{\cal O}_8$ is affected by QED
infrared divergences which are not relevant to our problem.}. We
will then consider at first only the operator $\hat{\cal O}_7$.

\section{Transverse momentum distribution in $b\rightarrow
s\gamma$}\label{calcolo}

The process we are dealing with has a very simple kinematics:
in lowest order it is the two-body decay $b \rightarrow
s\gamma$. Let us define
\begin{equation}\label{variabile}
x = {p^2_t \over m^2_b},
\end{equation}
where $p_t$ is the transverse momentum of the strange quark
with respect to the photon direction, fixed as $z$-axis, and $m_b$
is the mass of the heavy quark, to be identified with the hard
scale of the process\footnote{Let us note that $0\le x\le 1/4$.}.
\\
In lowest order the transverse momentum distribution then is
\begin{equation}
{d\Gamma \over dx} \ = \Gamma_0 \  \delta(x),
\end{equation}
that is the strange quark and the photon are emitted in
opposite directions, because of momentum conservation.
Acollinearity is generated by gluon emission; in
$b\rightarrow s\gamma g$, i.e. at $O(\alpha_S)$, $p_t =
-k_t$ while in $b\rightarrow s\gamma g_1\dots
g_n$, i.e. in higher orders,  $p_t = -k_{t 1} \dots -k_{t n}$.
\\
Beside the differential distribution the
partially integrated distribution\footnote{ Since we divide the
spectrum by the lowest-order rate $\Gamma_0$, we have that
$D(x=1/4)=\frac{\Gamma_{TOT}}{\Gamma_0}=1+O(\alpha_S)$. }
is also of interest
\begin{equation}\label{cumulativa}
D(x) = \int_0^x dx^\prime \ {1 \over \Gamma_0} \ {d\Gamma \over
dx^\prime}.
\end{equation}
Even though $\alpha_S(m_B)$ is small enough to justify a
perturbative approach, the combination $\alpha_S^n(m_B)\log^k x$,
with $0\le k \le 2n$ can be large. A resummation, to any order in $\alpha_S$, of
logarithms of the same magnitude is required to obtain
sensible physical results.
\\
A partial resummation of large logarithms with next-to-leading accuracy
has been performed in \cite{noi}: here we complete the calculation.
\\
It is well known, \cite{cttw}, that the resummation of large logarithms is
accomplished by an expression of the form:
\begin{equation}\label{master}
D(x) = K(\alpha_S) \Sigma(x;\alpha_S) + R(x;\alpha_S),
\end{equation}
where
\begin{itemize}
\item
$\Sigma(x;\alpha_S)$ is a universal, process-independent, function
resumming the infrared logarithms in exponentiated form.  It can be expanded in a series of functions as:
\begin{equation}\label{g}
\log \Sigma(x;\alpha_S)= L g_1(\alpha_S L) + g_2 (\alpha_S L)+
\alpha_S g_3(\alpha_S L)+\dots \ ,
\end{equation}
where $L=\log x$ (in general $L$ is a large infrared logarithm).
The functions $g_i$ have a power expansion of the form
\begin{equation}
g_i(z)=\sum_{k=0}^{\infty}g_{i,k}~z^k
\end{equation}
and resum logarithms
of the same size: in particular $g_1$ resums leading logarithms
of the form $\alpha_S^n~ L^{n+1}$ and $g_2$ the next-to-leading ones
$\alpha_S^n ~L^n$. The explicit form of  $\Sigma(x;\alpha_S)$ can be found in Ref.\cite{noi};
\item
$K(\alpha_S)$ is a short-distance coefficient function,
a process-dependent function, which can be calculated in perturbation theory:
\begin{equation}
K(\alpha_S) = 1+\frac{\alpha_S C_F}{\pi} k_1 + O\left(\alpha_S^2\right).
\end{equation}
\item
$R(x;\alpha_S)$ is the remainder function and satisfies
the condition
\begin{equation}\label{remainder}
R(x;\alpha_S) \rightarrow 0 \ \ \ {\rm{for}} \ \ \  x \rightarrow 0.
\end{equation}
It is process dependent, takes into account hard contributions and is calculable
as an ordinary $\alpha_S$ expansion:
\begin{equation}
R(x;\alpha_S)=\frac{\alpha_S C_F}{\pi} r_1(x)+ O\left(\alpha_S^2\right).
\end{equation}
\end{itemize}
The result is an improved perturbative distribution, reliable in
the semi-inclusive region \cite{cttw}, that is for small values of
$x$, which can be matched with a fixed-order spectrum, describing the
distribution for large values of $x$ \cite{librowebber}.
The description of the tools
used to perform the resummation of infrared logarithms is far from
the purpose of this note, and we refer the reader to the references
\cite{pp} -- \cite{cttw}.
\\
\\
In the next sections the full order $\alpha_S$ corrections for the coefficient
function $K(\alpha_S)$ and for the remainder function $R(x;\alpha_S)$ will be explicitely
calculated.

\section{$O(\alpha_S)$ corrections to $p_t$ distribution}
\begin{figure}[h]
\begin{center}
\mbox{\epsfig{file=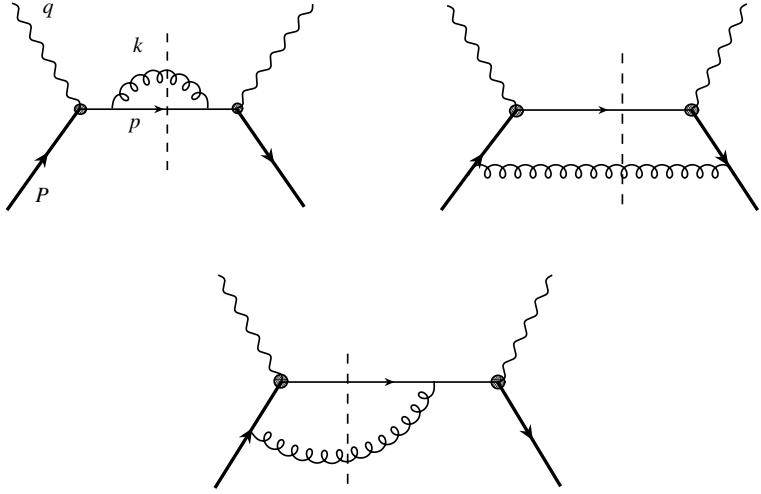, height=7cm}}
\caption{\label{reali}Real diagrams}
\end{center}
\end{figure}

In this section radiative corrections to the transverse momentum
distribution will be calculated e\-va\-lua\-ting the
Feynman diagrams depicted in fig. \ref{reali} for real gluon
emissions and in fig. \ref{virtuali} for virtual emissions.
We use the Feynman gauge where the gluon propagator is
\begin{equation}\label{feynprop}
D_{\mu\nu}(k)=\ -g_{\mu\nu}{i\over k^2 +i\epsilon}.
\end{equation}
The calculation is performed in dimensional regularization (DR) with
the dimension of the space-time
$$ n=4+\epsilon.$$ The operator $\hat{{\cal O}}_7$ from the basis
(\ref{base}) is inserted in the hard vertex, as discussed in
section \ref{sechamiltonian}.
Let us now define the kinematical variables: $P^\mu$ is the heavy
quark momentum, $p^\mu$ the light quark momentum, $k^\mu$ the
gluon momentum and $q^\mu$ the photon momentum: for real diagrams
it holds that $k^2=p^2=q^2=0$ and $P^2=m^2_b$, while, for virtual diagrams,
$k^2\neq 0$. The calculation is performed in the $b$ rest frame,
where
\begin{eqnarray}
P^\mu&=&(m_b,\vec{0})
\nonumber\\
q^\mu&=&(E_\gamma,0,0,E_\gamma).
\end{eqnarray}

\subsection{Real diagrams}
A straightforward evaluation of the diagrams in fig. \ref{reali} gives a contribution
to the rate:
\begin{equation}
\frac{d\Gamma}{\Gamma_0}={M(\omega,t;\epsilon)\over
\omega^{1-\epsilon}t^{1-\epsilon/2}}\  dt \ d\omega
= \left[{A_1(\omega,t;\epsilon)\over\omega^{1-\epsilon}t^{1-\epsilon/2}}+{S_1(t;\epsilon)\over\omega^{1-\epsilon}}+
{C_1(\omega;\epsilon)\over
t^{1-\epsilon/2}}+F_1(\omega,t;\epsilon)\right]\  dt \ d\omega \,
\end{equation}
where
\begin{eqnarray}\label{matrice}
\hskip 2cm A_{1} &\equiv &M(0,0;\epsilon)  \nonumber \\
S_{1}\left(
t\right) &\equiv &\frac{M(0,t;\epsilon)-M(0,0;\epsilon)}{t^{1-\epsilon/2}}  \nonumber \\
C_{1}\left( \omega \right) &\equiv &\frac{M(\omega
,0;\epsilon)-M(0,0;\epsilon)}{\omega^{1-\epsilon} }  \nonumber \\
F_{1}\left( \omega ,t;\epsilon\right) &\equiv &\frac{M(\omega
,t;\epsilon)-M(0,t;\epsilon)-M(\omega
,0;\epsilon)+M(0,0;\epsilon)}{\omega^{1-\epsilon}
\,t^{1-\epsilon/2}}
\end{eqnarray}
and
$$
\omega={2P\cdot k \over m_b^2}=\frac{2E_g}{m_b},  ~~~~~~ t={1-\cos\theta \over 2},
$$
with $\theta$ the angle between the gluon and the direction
$-\hat{z}\;$\footnote{Let us remember that the direction $+\hat{z}$ is fixed by the
photon space momentum.}.
\\
It follows from their definition that the functions
$A_1(\omega,t;\epsilon)$, $S_1(\omega;\epsilon)$, $C_1(t;\epsilon)$
and $F_1(\omega,t;\epsilon)$ are finite in the soft and the collinear limit,
defined respectively as
\begin{equation}\label{irlimite}
\omega \rightarrow 0   \;\;\;\;\;{\rm and}\;\;\;\; t\rightarrow 0.
\end{equation}
The rate in eq. (\ref{matrice}) has to be integrated over
the whole phase space with the kinematical constraint
\begin{equation}\label{vincolo}
\delta[x-\omega^2t(1-t)],
\end{equation}
which selects gluons with transverse momentum $p_t^2=x m_b^2$
\footnote{Let us
recall that for the single gluon emission
$x\equiv p^2_t /m^2_b = k^2_t / m^2_b$.}.
The cumulative distribution takes a contribution of the form:
\begin{equation}
D_R(x)=\int_0^xdx^\prime \int_0^1 d\omega \int_0^1 dt \ {1 \over
\Gamma_0} \ {d\Gamma \over dx^\prime}(\omega,t;\epsilon) \
\delta[x^\prime-\omega^2t(1-t)].
\end{equation}
After the integration we expect four kinds of terms:
\begin{itemize}
\item
Poles in the regulator $\epsilon$: they parametrize the infrared
singularities and cancel in the sum with virtual diagrams because
the distribution we are dealing with is infrared-safe\footnote{
A distribution is infrared-safe if it is insensitive to the emission
of a soft and a collinear gluon \cite{librowebber}.};
\item
Logarithmic terms diverging for $x\rightarrow 0$;
\item
Constant terms: they enter the coefficient function
$K(\alpha_S)$;
\item
Remainder functions: terms that vanish in the limit $x\rightarrow 0$.
\end{itemize}
Integrating over $x^\prime$ we have:
\begin{equation}
D_R(x)= \int_0^1 d\omega \int_0^1 dt \ {1 \over
\Gamma_0} \ {d\Gamma \over dx^\prime}(\omega,t;\epsilon) \
\theta[x-\omega^2t(1-t)].
\end{equation}
The remaining integrations are non-trivial because of the simultaneous presence of
the kinematical constraint and by the dimensional regularization parameter $\epsilon$.
By using the identity
\begin{equation}
\theta[x-\omega^2t(1-t)]~=~1-\theta[\omega^2t(1-t)-x],
\end{equation}
we separate these two effects and rewrite the
distribution $D_R(x)$ as a difference between an
integral over the whole phase space
and a integral over the complementary region:
\begin{equation}
D_R(x)=\int_0^1 d\omega \int_0^1 dt{1 \over \Gamma_0} \ {d\Gamma
\over dx}(\omega,t;\epsilon) \ - \ \int_0^1 d\omega \int_0^1 dt {1
\over \Gamma_0} \ {d\Gamma \over dx}(\omega,t;0) \
\theta[\omega^2t(1-t)-x]~+~O(\epsilon).
\end{equation}
The first integral must be evaluated for $\epsilon\neq 0$ because it contains
poles in $\epsilon$, but is done over a very simple domain, independent of $x$.
The second integral does not contain any pole in $\epsilon$
and therefore one can take the limit $\epsilon\rightarrow 0$ in the integrand.
It depends on the kinematics of the
process and can be integrated by introducing a suitable basis of
harmonic polylogarithms as in \cite{remiddi}. The most convenient
basis we found  consists of the basic functions
\begin{eqnarray}\label{baseint}
g[0;y]&\equiv& {1 \over y} \nonumber \\ g[-1;y]&\equiv& {1 \over
y+1} \nonumber \\ g[-2;y]&\equiv& {1 \over {\sqrt{y}(1+y)}}\nonumber \\
g[-3;y]&\equiv& -{\sqrt{x} \over {2({1-\sqrt{x}\sqrt{y})}\sqrt{y}}}.
\end{eqnarray}
The harmonic polylogarithms (HPL) of weight 1 are defined as:
\begin{eqnarray}
J[a;y]&\equiv&\int_0^y dy^\prime \ g(a;y^\prime)  \ \ \ \rm{for}\;   a \neq 0 \nonumber\\
J[0;y]&\equiv& \log y.
\end{eqnarray}
In terms of usual functions, they read:
\begin{eqnarray}
J[-1;y]&\equiv& \log (1+y)\nonumber\\
J[-2;y]&\equiv& 2\arctan \sqrt y\nonumber\\
J[-3;y]&\equiv& \log (1-\sqrt{x} \sqrt{y})
\end{eqnarray}
The HPL's of weight 2 are defined for $(u,v)\not= (0,0)$ as
\begin{equation}
J[u,v;y]\equiv \int_0^y dy^\prime \ g[u;y^\prime]\int_0^{y^\prime}
 dy^{\prime\prime} g[v;y^{\prime\prime}]
\end{equation}
and $J[0,0;y]=1/2\log^2y$.
HPL s of higher weight may be defined in an analogous way. They will not be used here.
\\
The final result for real diagrams turns out to be
\begin{equation}
D_R(x)= C_F {\alpha_S \over \pi}\ \left({m^2_b \over
4\pi\mu^2}\right)^{\epsilon/2}\ {1\over \Gamma(1+\epsilon/2)}\ \left[{2\over
\epsilon^2} \ - {5 \over 2\epsilon}\ -{1 \over 4} \ \log^2 x - \
{5 \over 4} \ \log x + \ {1 \over 4} \ +d(x)\right],
\end{equation}
where $d(x)$ is a function vanishing for $x\rightarrow 0$.
\\
The matrix elements of the remaining operators $\hat{\mathcal{O}}_i\; (i\not= 7)$
do not contain infrared divergences.
Therefore their contributions to $D_R$ do not involve (infrared)
poles in $\epsilon$, logarithms of $x$ and constants,
but only new functions, which vanish in the limit $x\rightarrow 0$.

\subsection{Virtual diagrams}
\begin{figure}[h]
\begin{center}
\mbox{\epsfig{file=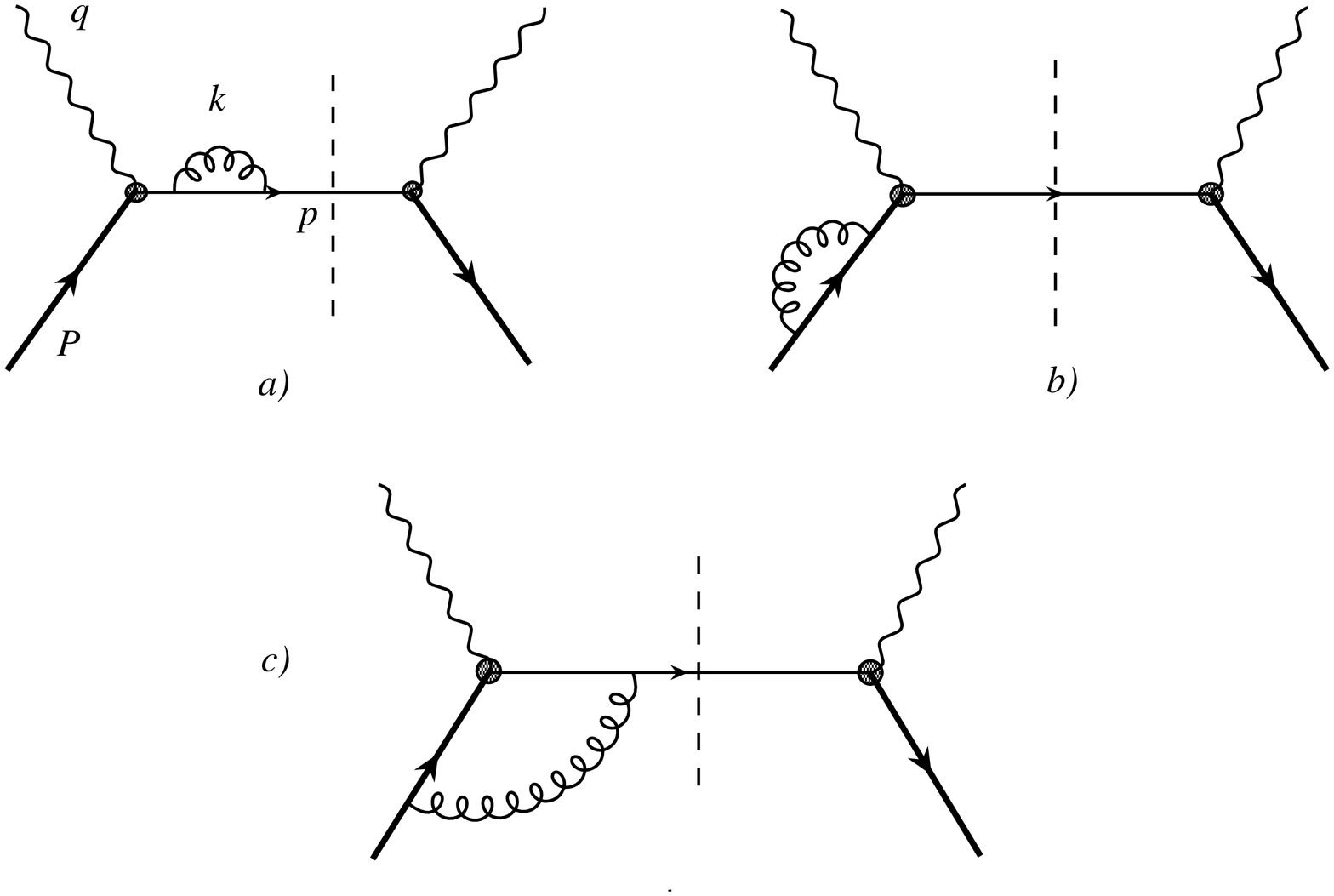, height=7cm}}
\caption{\label{virtuali}Virtual diagrams}
\end{center}
\end{figure}

Virtual corrections to $b\rightarrow s\gamma$ have been calculated
in \cite{greubetal,pott} for a massive strange quark; we present
here the computation in the massless case.
The diagrams consist of self-energy corrections to the heavy and light
lines and of vertex corrections to the operator $\hat{\mathcal{O}}_7$
(see fig. \ref{virtuali});
we compute them in the $\overline{MS}$ scheme so as to be consistent with the (known) coefficient functions $C_i$.
The computation can be done with standard Feynman parameter technique
or by a reduction using the integration by part identities \cite{chettak}.
Let us briefly describe the evaluation of the vertex correction
within the latter method. One has to compute the scalar integral:
\begin{equation}
\mathcal{V}=\int {d^nk \over (2\pi)^n} \ {N(k^2,P\cdot k,
p\cdot k;\epsilon)\over k^2 [(k-P)^2-m^2](k-p)^2},
\end{equation}
where
\begin{equation}
N(k^2,P\cdot k, p\cdot k;\epsilon) = 32 P\cdot k -32 p\cdot k - 16 m^2_b
+\mathcal{O}(\epsilon^2) k^2 +
\mathcal{O}(\epsilon^2)  P\cdot k p\cdot k +
\mathcal{O}(\epsilon^2)  {(p\cdot k)}^2  .
\end{equation}
$\mathcal{V}$ has at most a double pole in $\epsilon$ coming from
the product of the soft and the collinear singularities.
The terms in the numerator $N$, which vanish in the soft limit
$k_{\mu}\rightarrow 0$, do not give rise to soft singularities
and therefore produce at most a simple pole
coming from the collinear or the ultraviolet region.
Therefore the $\mathcal{O}(\epsilon^2)$ terms in $N$
do not contribute in the limit $\epsilon\rightarrow 0$.
\\
By expressing the scalar products in the numerator as linear combinations
of the denominators as
\begin{eqnarray}
\label{rotate}
k\cdot p &=& \frac{1}{2} \left( k^2 - (k-p)^2 \right),
\nonumber \\
k\cdot P &=& \frac{1}{2} \left( k^2 - (k-P)^2 + m_b^2\right),
\end{eqnarray}
we can reduce $\mathcal{V}$ to a superposition
of scalar integrals of the form:
\begin{equation}
{\rm T}[a,b,c]=\int {d^nk \over (2\pi)^n} \ {1 \over [k^2]^a \
[(k-P)^2-m_b^2]^b \ [(k-p)^2]^c}
\end{equation}
with $a,b,c\le 1$.
The above amplitudes can be related to each other
by identities of the form \cite{chettak}:
\begin{equation}
\int d^n k \ {\partial \over \partial k^\mu} \  {  v^{\mu}
 \over [k^2]^a [(k-P)^2-m^2]^b[(k-p)^2]^c  } =0
\end{equation}
with $v^{\mu}=k^\mu,p^\mu,P^\mu$.
By explicitly evaluating the derivatives and re-expressing the scalar
products using eqs. (\ref{rotate}), one obtains relations among amplitudes
with shifted indices.
\\
By solving the above identities, one can reduce all the amplitudes
to the tadpole, and one obtains for our integral\footnote
{Such a strong reduction of 3-point function to a vacuum amplitude
is possible because the only scale in the process is the heavy quark mass $m_b$.
Virtual corrections have indeed the lowest-order
kinematics $P^2 = m_b^2,~P\cdot p = m_b^2/2,~p^2 = q^2 = 0$.}:
\begin{equation}
\mathcal{V} = \left(-{16 \over \epsilon} +8-8\epsilon\right) {1\over
m^2_b} \ {\rm T}[0,1,0],
\end{equation}
where
\begin{equation}
{\rm T}[0,1,0]= C_F{\alpha_S\over 4\pi} \ \left({m^2\over
4\pi\mu^2}\right)^{\epsilon / 2} \ {\Gamma(-\epsilon / 2)\over
1+\epsilon/2} \ m_b^2.
\end{equation}
Summing self-energies and vertex corrections,
and subtracting the $1/\epsilon$ poles according to the $\overline{MS}$ scheme,
one obtains for their contribution to the rate $D_V$
\footnote{To factorize $\Gamma_0$ one has to replace
$m_{b,\overline{MS}}(\mu_b)$ by $m_{b,\overline{MS}}(m_b)$
using the formula
$m_{b,\overline{MS}}(\mu_b)=m_{b,\overline{MS}}(m_b)
( 1+\frac{3}{2} {C_F\alpha_S\over\pi})$.}:
\begin{equation}
D_V=C_F{\alpha_S\over\pi} \ \left({m^2_b\over 4\pi\mu^2}\right)^{\epsilon/2}
\ \Gamma\left(1-{\epsilon\over 2}\right) \ \left[-{2\over \epsilon^2}+{5 \over
2\epsilon} + 4 \log{m_b\over \mu_b} -3\right].
\end{equation}
We have kept the factor in front of the square bracket unexpanded
to simplify the computation of the total rate.
\\
The virtual corrections to the remaining operators $\hat{\mathcal{O}}_{i\not= 7}$
contain only (simple) ultraviolet poles in $\epsilon$,
which are removed by renormalization; their contributions to $D_V$
amount only to finite constants and $\log m_b/\mu_b$.

\subsection{Final result}
Summing real and virtual contributions, the transverse momentum
distribution for the decay $b\rightarrow s\gamma$ reads, to
$O(\alpha_S)$:
\begin{equation}\label{final}
D(x)=1~+~C_F{\alpha_S\over\pi}\ \left[-{1 \over 4} \ \log^2 x - \ {5 \over
4} \ \log x + f +\ d(x)\right].
\end{equation}
As expected, the result contains a double logarithm
and a single logarithm of $x$, a finite term $f$ and a function $d(x)$
vanishing in the limit $x\rightarrow 0$.
\\
By expanding the resummed formula to order $\alpha_S$ one obtains:
\begin{eqnarray}\label{final2}
D(x) &=& \left( 1+\frac{C_F\alpha_S}{\pi} k_1 \right)
\left(1-\frac{A_1}{4}\alpha_S \log^2 x + B_1 \alpha_S \log x \right)+
\frac{C_F\alpha_S}{\pi} r(x)
\nonumber
\\
&=& 1-\frac{A_1}{4}\alpha_S \log^2 x + B_1 \alpha_S \log x+\frac{C_F\alpha_S}{\pi} k_1
+\frac{C_F\alpha_S}{\pi} r(x)+O(\alpha_S^2).
\end{eqnarray}
By identifying the resummed result expanded to $O(\alpha_S)$ with the fixed-order one
--- matching procedure ---
we check the
values for $A_1$ and $B_1$ evaluated in our previous paper using
general properties of QCD radiation \cite{noi}
and we extract the value of the coefficient function:
\begin{equation}\label{coefficient}
k_1~=~f~=~ - \ {11 \over 4}- \ {\pi^2 \over 12} +~ 4\log{m_b\over \mu},
\end{equation}
as well as the remainder function $r_1(x)=d(x)$.
As explained in previous sections,
the remaining operators $\hat{\mathcal{O}}_{i\not= 7}$ contribute to $D(x)$ only
by finite terms $\tilde{r}_i$ and remainder functions.
Since the constants $\tilde{r}_i$ come from virtual diagrams alone, we can quote their
result from \cite{greubetal} and present an improved formula for the
coefficient function, in analogy with \cite{acg}:
\begin{equation}\label{coefcompleto}
K(\alpha_S)~=~ 1 + {\alpha_S \over 2\pi}
\sum_{i=1}^8 {C^{(0)}_i(\mu_b) \over C^{(0)}_7(\mu_b)}\left(\Re\  \tilde{r}_i +
\gamma^{(0)}_{i7}\log{m_b\over \mu_b}\right)+{\alpha_S \over
2\pi}{C^{(1)}_7(\mu_b) \over
C^{(0)}_7(\mu_b)}+\mathcal{O}(\alpha_S^2)
\end{equation}
where
\begin{eqnarray}
\tilde{r}_{i} &=& r_i ~~~~~ i\not= 7
\nonumber \\
\tilde{r}_7 &=& \frac{8}{3}\left( f-4\log\frac{m_b}{\mu_b}\right)=
-\frac{22}{3}-\frac{2\pi^2}{9}.
\end{eqnarray}
Let us remark that only the coefficients related to the operators
with $i=1,2,7,8$ are
relevant, because the others are multiplied by very small
coefficient  functions and can be neglected:
\begin{eqnarray}
r_1&=&-{1 \over 6}r_2\nonumber\\
\Re \ r_2&=& -4.092-12.78(0.29-m_c/m_b)\nonumber\\
r_8&=& {4 \over 27}(33-2\pi^2).
\end{eqnarray}
The analytic expressions for the coefficient functions as well as a standard
numerical evalutation are given in \cite{misiak2}.
The anomalous dimension $\gamma^{(0)}_{77}$ is derived from the coefficient
of the logarithmic term in $k_1$.
The values of $\gamma_{i7}^{(0)}$ are \cite{misiak2}:
\begin{equation}
\gamma^{(0)}_{i7}=\left(-{208 \over 243}, {416 \over 81}, -{176 \over
81}, -{152 \over 243}, -{6272\over 81}, {4624 \over 243},{32 \over
3},-{32 \over 9}\right).
\end{equation}
\\
Equation(\ref{coefcompleto}) is the main result of our paper and allows a
complete resummation to NLO of transverse momentum logarithms.
\\
The explicit calculation of the remainder function in (\ref{final2}) reads
\begin{eqnarray}\label{taudef}
&r(\tau)&=\frac{(\tau-1)(49\tau^8+468\tau^7+1797\tau^6+3642\tau^5+4450\tau^4+3642\tau^3+1797\tau^2+468\tau+49)}{12(\tau+1)^5(\tau^2+3\tau+1)^2}
\nonumber\\
&+&
\frac{-5-61\tau-317\tau^2-912\tau^3-1622\tau^4-1934\tau^5-1622\tau^6-912\tau^7-317\tau^8-61\tau^9-5\tau^{10}}{4(\tau+1)^6(\tau^2+3\tau+1)^2}\log\tau
\nonumber\\
&-&J[0,-3,\tau]+J[0,-3,1/\tau]-2J[0,-1,\tau]+J[-1,0,\tau]+J[-1,-3,\tau]-J[-1,-3,1/\tau]
\nonumber\\
&-&2\sqrt \tau \arctan(\sqrt\tau)\frac{(\tau+1)(2\tau^2+7\tau+2)}{(\tau^2+3\tau+1)^2}+\frac{\pi}{2}\sqrt
\tau\frac{(\tau+1)(2\tau^2+7\tau+2)}{(\tau^2+3\tau+1)^2}+\frac{49}{12}
\nonumber\\
&+&\frac{5}{4}\log\tau-\frac{5}{2}\log(\tau+1)+\log^2(\tau+1),
\end{eqnarray}
where
\begin{equation} \label{tau}
\tau=\frac{1-\sqrt{1-4x}}{1+\sqrt{1-4x}}.
\end{equation}
Let us notice that $\tau$ behaves as $x$ for small
values of the transverse momentum
\begin{equation}
\tau(x)=x+O(x^2)
\end{equation}
and it is a unitary variable
\begin{eqnarray*}
\tau\rightarrow 0 \ \ \ &\rm{for}& \ \ \ x\rightarrow 0\\
\tau\rightarrow 1 \ \ \ &\rm{for}& \ \ \ x\rightarrow 1/4.\\
\end{eqnarray*}
The relation (\ref{tau}) may be inverted as
\begin{equation}
x=\frac{\tau}{(\tau+1)^2}.
\end{equation}
One can easily check that $r(\tau)$ vanishes for $\tau\sim
x\rightarrow 0$, by using the properties
\begin{equation}
J[0,-1,0]=J[0,-3,0]=J[-1,0,0]=J[-1,-3,0]=0
\end{equation}
\begin{equation}
\lim_{\tau\to 0}J[0,-3,1/\tau]=\lim_{\tau\to
0}J[-1,-3,1/\tau]=-\frac{\pi^2}{3}.
\end{equation}

\section{Conclusions}
To sum up: eq. (\ref{final}), which contains the final result, represents the full
evaluation to $O(\alpha_S)$ of the transverse momentum
distribution. It is explicitly given in terms of an analytic expression.
\\
Contrary to what happens in hard processes at much larger
energies, at the energy scales involved here for the $b$ decay the
remainder function contribution does play a more important role.
\\
A straightforward numerical evaluation of the remainder function $r(x)$ of
eq. (\ref{final2}) allows us to conclude that its contribution can be
safely neglected for small values of $x$, up to $x\simeq 0.1$,
where it approaches the zero limit of $x\rightarrow 0$.
For larger values of $x$, however, the size of its contribution increases
to reach values of the order of the $10$--$15\%$ of the combined
leading and next-to-leading logarithmic terms.
\\
A detailed report describing the calculation giving rise the $O(\alpha_S)$
evaluation presented here, together
with an analysis of the related phenomenological impact, will be presented in
a future article \cite{noi1}.

\end{document}